\documentclass{PoS}
\usepackage{bm}       
\usepackage{amsmath}
\usepackage{cancel}

\newcommand{\be}{\ensuremath{\beta} }
\newcommand{\cO}{\ensuremath{\mathcal O} }
\newcommand{\X}{\ensuremath{\!\times\!} }
\newcommand{\lsim}{\ensuremath{\lesssim} }
\newcommand{\Sb}{\ensuremath{\cancel{S^4}} }
\newcommand{\refcite}[1]{Ref.~\cite{#1}}
\newcommand{\eq}[1]{Eq.~\ref{#1}}
\newcommand{\fig}[1]{Fig.~\ref{#1}}

\title{Finite size scaling and the effect of the gauge coupling in 12 flavor systems}
\ShortTitle{Finite size scaling and the effect of the gauge coupling in 12 flavor systems}

\author{\speaker{Anna Hasenfratz}, Anqi Cheng, Gregory Petropoulos, David Schaich\footnote{Present address: Department of Physics, Syracuse University, Syracuse, NY 13244} \\
        Department of Physics, University of Colorado, Boulder, CO-80309-390\\
        E-mail: \email{anna@eotvos.colorado.edu}}

\abstract{Finite size scaling is a powerful tool to study the critical properties of systems governed by one relevant operator, assuming all irrelevant operators have scaling dimensions much smaller then zero.  This condition is likely not satisfied in many-fermion conformal systems where perturbation theory predicts a nearly-marginal irrelevant gauge coupling.  In this work we carry out a new investigation of SU(3) lattice gauge theory with 12 fundamental flavors.  Analyzing data at many different gauge couplings, our preliminary results indicate that a finite size scaling analysis that takes into account the effect of a nearly-marginal gauge coupling can resolve many of the inconsistencies observed previously in this system, leading to results consistent with conformal infrared dynamics and predicting a mass scaling anomalous around $\gamma_m=0.25$.}

\FullConference{31st International Symposium on Lattice Field Theory - LATTICE 2013\\
                July 29 - August 3, 2013\\
                Mainz, Germany}

\begin{document}
\section{Introduction}
Asymptotically free gauge theories with many fermionic degrees of freedom can exhibit unusual infrared properties at strong gauge coupling.
Some develop a new conformal fixed point with possibly large anomalous dimensions.
Others remain chirally broken but their dynamics may be approximately scale-invariant across a wide range of energies and  could be candidates for Beyond-Standard Model physics.
In either case there is interesting non-perturbative infrared dynamics worth studying.
Lattice gauge calculations are particularly suitable to investigate these strongly-coupled systems, and in recent years significant computational resources have been devoted to this effort.

The SU(3) gauge model with $N_f=12$ fundamental fermions is a  controversial system.
Several groups have studied the infrared properties of this model using different methods and different lattice actions, arriving at contradictory conclusions regarding its IR dynamics. 
(For a limited set of references see Refs.~\cite{Appelquist:2009ty, Deuzeman:2009mh, Hasenfratz:2011xn, Fodor:2011tu, DeGrand:2011cu, Cheng:2011ic, Cheng:2013eu, Fodor:2012uw, Fodor:2012et, Appelquist:2011dp, Aoki:2012eq, Aoki:2013pca, Itou:2012qn, Lin:2012iw, Jin:2012dw} and the recent review \refcite{Giedt:2012LAT}.)
Finite size scaling is one of the methods frequently used to study this system~\cite{DeGrand:2011cu, Fodor:2012et, Aoki:2012eq}. 
While 12-flavor data appear to obey finite size scaling, recent studies find that different physical quantities predict inconsistent scaling exponents, which suggests that it is not possible to consistently describe all the data assuming conformal dynamics~\cite{Fodor:2012et, Aoki:2012eq}.

Finite size scaling techniques provide an effective tool to investigate models governed by a fixed point with only one relevant operator, especially if the irrelevant operators are strongly irrelevant, i.e., their scaling dimensions are much below zero.
If this condition is not met, either very large volumes have to be used, or corrections to scaling have to be taken into account.
Both perturbation theory and non-perturbative step scaling function calculations predict that in the 12-flavor systems the gauge coupling has very small scaling exponent, $-0.3 \lsim y_0 \lsim -0.1$~\cite{Ryttov:2010iz, Appelquist:2009ty}.
In this paper we consider the possibility that some of the inconsistencies found in earlier investigations are due to this nearly-marginal gauge coupling.

\begin{figure}[b]
\centering
  \includegraphics[width=0.45\linewidth]{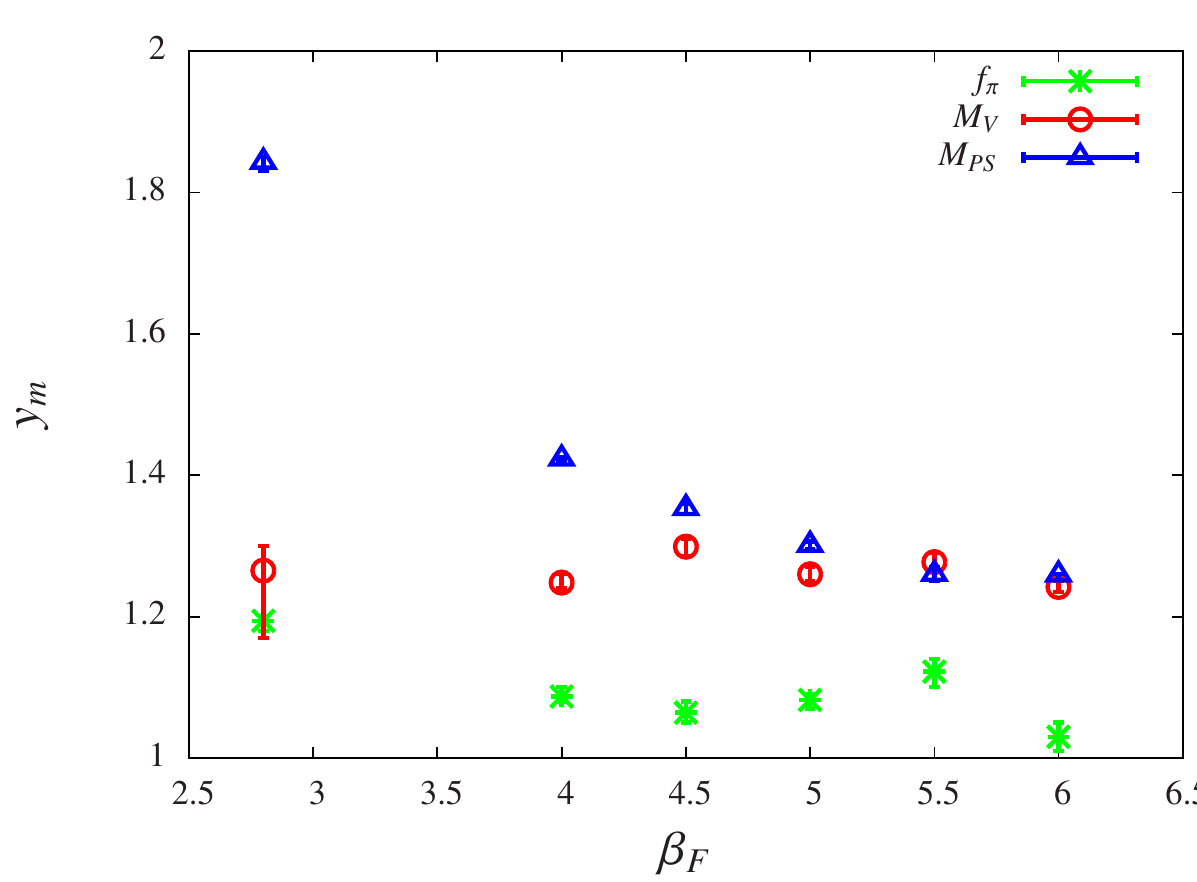}\hfill
  \includegraphics[width=0.45\linewidth]{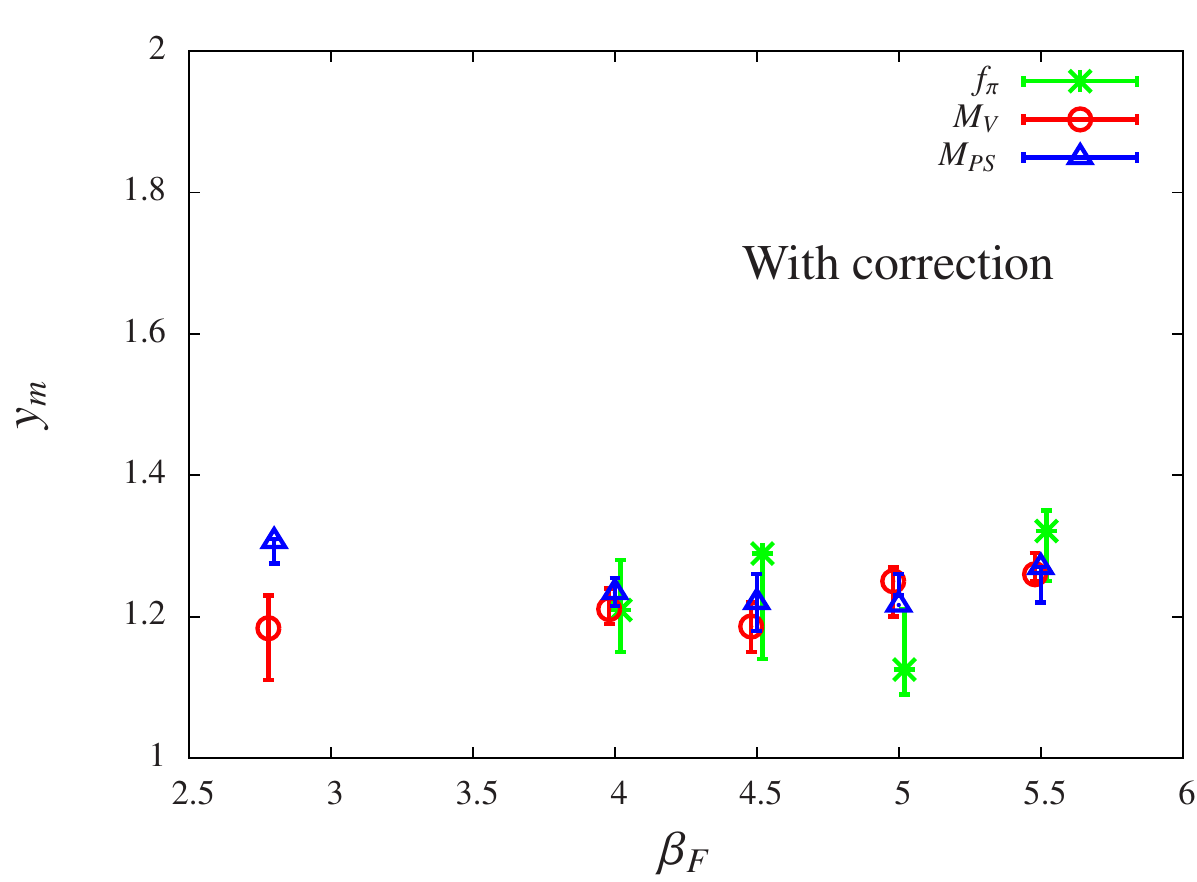}
  \caption{\label{fig:scaling_exp} The scaling dimension $y_m$ predicted by finite size scaling, as a function of the gauge coupling $\beta_F$ for the pseudoscalar (blue triangles), vector (red circles) and $f_\pi$ (green $\times$s).  Left: fits including only the relevant mass operator.  Right: fits including both the relevant operator and leading irrelevant corrections.}
\end{figure}

In order to investigate the effects of a nearly-marginal irrelevant gauge operator, it is essential to study the system at many gauge coupling values.
In this work we cover a wide range from a strong coupling near the onset of the ``$\Sb$'' lattice phase~\cite{Cheng:2011ic} to as weak coupling as our lattice volumes allow.
We find that finite size scaling using only the leading relevant operator predicts scaling exponents that depend both on the physical quantity considered as well as on the bare gauge coupling, as shown in the left panel of \fig{fig:scaling_exp}.
When we include the corrections to scaling due to the nearly-marginal gauge coupling, our preliminary analysis  predicts scaling exponents that are, within errors, independent of the gauge coupling and consistent for the pseudoscalar meson, vector meson, and $f_\pi$, as shown in the right panel of \fig{fig:scaling_exp}.

While we cannot prove that all physical quantities will scale consistently once corrections to scaling are taken into account -- especially because these corrections might be more important to some quantities than to others -- our results resolve some of the existing controversies of the 12-flavor system and reinforce the IR-conformal interpretation suggested by our earlier studies of the bare step scaling function~\cite{Hasenfratz:2011xn}, phase transitions~\cite{Hasenfratz:2013uha} and Dirac eigenvalues~\cite{Cheng:2013eu}.
Our finite size scaling results prefer a fairly small anomalous dimension, $\gamma_m^{\star} = y_m^{\star} - 1 \approx 0.25$.
The statistical errors on $\gamma_m^{\star}$ are about 10\%, with similar systematic uncertainties for the three quantities considered.
At this point we cannot give a more precise error estimate, but note that this value is consistent with our findings for $\gamma_m^{\star}$ from the Dirac operator spectral density~\cite{Cheng:2013eu}.

\section{Numerical setup}
In our numerical studies we use nHYP smeared staggered fermions with smearing parameters $(0.5,0.5,0.4)$ to ensure the numerical stability of simulations.
Our gauge action contains fundamental and adjoint plaquette terms with $\beta_A/\beta_F=-0.25$ to avoid the potential scaling violation effects known to exist at positive adjoint plaquette coupling.
In \refcite{Cheng:2011ic} we reported on the phase structure and other properties of this action.

In our previous studies we were able to run simulations in the $m=0$ chiral limit with periodic spatial boundary conditions on volumes as large as $32^3\X64$ at gauge couplings up to and within the \Sb phase~\cite{Hasenfratz:2013uha}.
In the present work we consider gauge couplings $\beta_F=2.8$, 4.0, 4.5, 5.0, 5.5 and 6.0 on volumes $16^3\X32$, $20^3\X40$, $24^3\X48$ and $32^3\X64$.
At the strongest gauge couplings we can also use $12^3\X24$ volumes.
We choose the bare mass in the range $0.005 \leq m \leq 0.12$, requiring that the vector meson mass $M_V \lsim 0.8$.

\begin{figure}[tb]
\centering
  \includegraphics[width=0.45\linewidth]{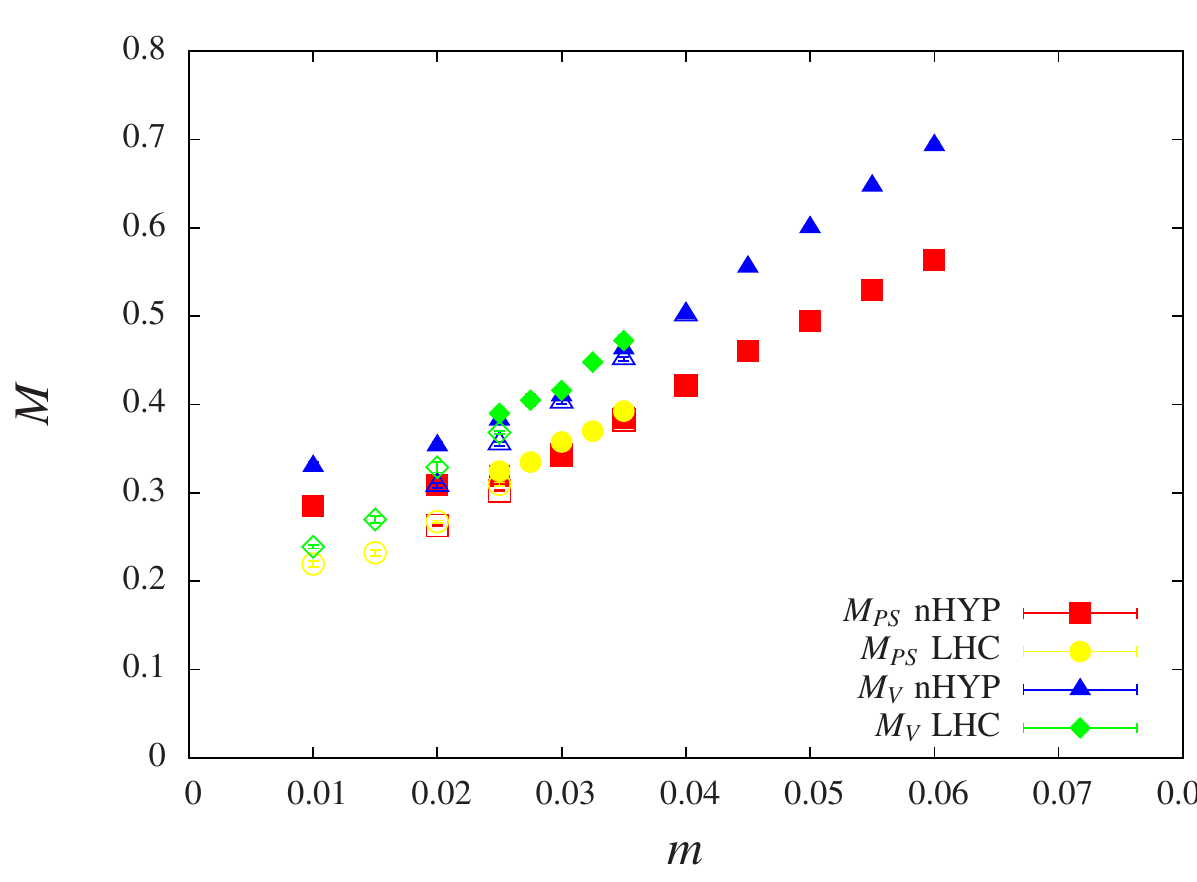}\hfill
  \includegraphics[width=0.45\linewidth]{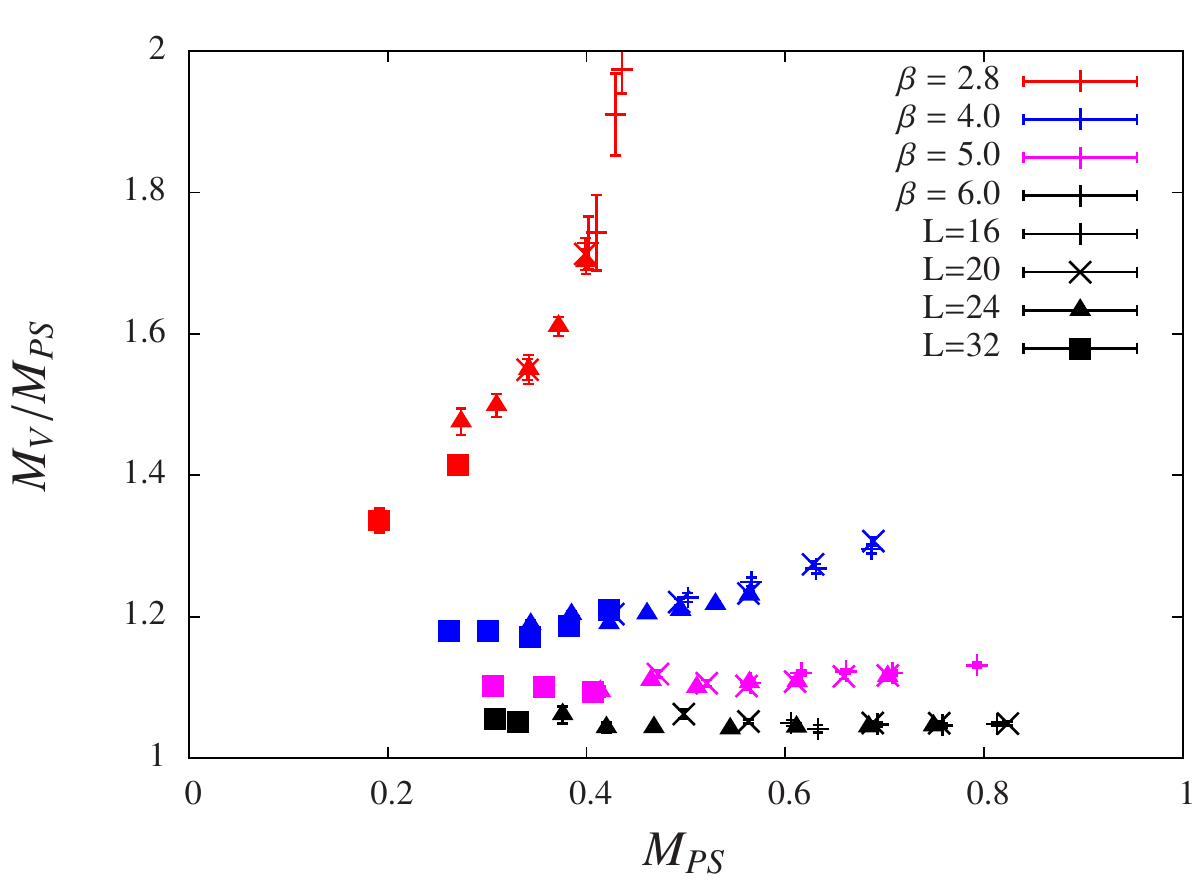}
  \caption{\label{fig:compare} Left: The pseudoscalar and vector meson masses as the function of the bare fermion mass for our nHYP action at $\beta_F=4.0$, and for the LHC action at $\beta=2.2$, from \refcite{Fodor:2011tu}. Only $24^3\X48$ (filled symbols) and $32^3\X64$ (open symbols) data are included.    Right: Ratio of vector and pseudoscalar meson masses as the function of the pseudoscalar mass at various gauge coupling values (nHYP data).} 
\end{figure}

It is instructive to compare our spectrum data with the results published by the Lattice Higgs Collaboration (LHC) in \refcite{Fodor:2011tu}.
We found (rather accidentally) that $\beta_F=4.0$ in our action matches the LHC $\be = 2.2$ stout-smeared spectrum very closely, as illustrated in the left panel of \fig{fig:compare}.
The agreement of both the pseudoscalar and vector meson spectrum as functions of the bare mass suggests that the mass renormalization factors of the two actions are nearly identical.
This is not very surprising, given that both actions employ smeared staggered fermions.
We have not been able to find similar  match with the results of the LatKMI Collaboration~\cite{Aoki:2012eq}, who use the Highly Improved Staggered Quark action.
The best estimate we can make is that our $\beta_F=5.5$, 6.0 couplings are close to the $\beta=3.7$, 4.0 of \refcite{Aoki:2012eq}, but with different mass renormalization factors.

In the right panel of \fig{fig:compare} we show the dimensionless ratio $M_V/M_{\pi}$ at different gauge coupling values, as functions of the pseudoscalar mass.
In a chirally broken system this ratio diverges as $1/M_\pi$ in the chiral limit and, at least in the scaling regime, different $\beta$ values can be rescaled with the lattice spacing $a$ to form a unique curve.
In a conformal system the ratio should approach a constant value in the chiral limit.
At finite mass different $\beta$ values could predict different ratios due to corrections from the irrelevant operators.

Our data show no indication of divergence of $M_V / M_{\pi}$ in the chiral limit; in fact at stronger gauge couplings the ratio monotonically decreases with decreasing $M_{\pi}$.
If the $N_f=12$ flavor system were chirally broken, this indicates that none of our calculations are close enough to the chiral limit to probe spontaneous chiral symmetry breaking.
The conformal scenario is more plausible, as the different gauge coupling data could be consistent with a ratio $M_V/M_{\pi}\approx 1.1$ in the chiral limit.
We observe very strong mass dependence at the strongest coupling, $\beta_F=2.8$, signaling large cut-off effects.
This is not surprising as this coupling is near the edge of the \Sb phase.
The three other data sets at $\beta_F=4.0$, 5.0 and 6.0 show considerably weaker mass dependence.
At $\beta_F=4.0$ the ratio still decreases as $M_{\pi}$ decreases, at $\beta_F=5.0$ it is approximately constant, and at $\beta_F=6.0$ we see a slight increase towards the chiral limit.
This qualitative change might indicate a conformal fixed point around $\beta_F^{\star} \sim 5.0$, in the scheme defined by this observable.

\section{Finite size scaling}
Finite size scaling is a well understood technique in statistical physics.
Its derivation is easiest using renormalization group arguments and has been reviewed recently in connection with infrared conformal systems \cite{DeGrand:2009mt, DelDebbio:2010ze}.
Here we summarize only the steps relevant for the scaling of physical quantities ``$M_H$'' with mass (engineering) dimension $[M_H] = 1$.

For concreteness consider a system with one relevant operator, denoted by $m$, that has a scaling dimension $y_m > 0$.
All other operators, denoted by $g_i$, are irrelevant with scaling exponents $y_i < 0$.
Renormalization group arguments predict that in a finite spatial volume $L^3$, $M_H$ depends only on specific combinations of the couplings, and can be written as
\begin{equation}
  \label{eq:general}
  M_H = L^{-1} f\left(x, g_i m^{-y_i / y_m}\right),
\end{equation}
where $x \equiv L m^{1 / y_m}$.
In the critical $m\to 0$ limit, $g_i m^{-y_i / y_m} \to 0$ and we find the familiar finite-size scaling formula
\begin{equation}
\label{eq:leading_fss}
  M_H = L^{-1} f(x),
\end{equation}
where $f(x)$ is an arbitrary but unique scaling function.
It is important to note that the scaling function $f(x)$ depends on the observable $M_H$, but the exponent $y_m$ in the scaling variable $x$ is universal, characteristic of the corresponding fixed point.

If one of the irrelevant operators, let's say $g_0$, is nearly marginal with scaling exponent $y_0 \lsim 0$, the term $g_0 m^{-y_0 / y_m}$ can remain significant and has to be included in the scaling analysis.
This leads to the modified finite-size scaling formula
\begin{equation}
  M_H = L^{-1} f\left(x, g_0 m^{\omega}\right),
\end{equation}
where $\omega \equiv -y_0 / y_m \gtrsim 0$.
The scaling function $f\left(x, g_0 m^{\omega}\right)$ is analytic even at the fixed point, and can be expanded as
\begin{equation}
  \label{eq:expansion}
  L M_H = F(x)\left\{1 + g_0 m^{\omega} G(x) + \cO\left(g_0^2 m^{2\omega}\right)\right\}.
\end{equation}
The first term is the usual finite-size scaling expression; the second term accounts for leading corrections to scaling due to the nearly-marginal gauge coupling.

In the limit $x \to 0$, both $F(x)$ and $G(x)$ approach finite constants.
In the infinite-volume limit, with small but fixed $m$, $F(x)\propto x$ while $G(x)$ remains finite.
Our simulations cover a limited range $0.5 \lsim x \lsim 5$, over which we approximate $G(x)$ by a constant, $G(x) = c_G$, so that 
\begin{equation}
  \label{eq:corrected}
  \frac{L M_H}{1 + c_G g_0 m^{\omega}} = F(x).
\end{equation}
We test the validity of this assumption by repeating our analyses using subsets of our data restricted to smaller ranges in $x$.
\eq{eq:corrected} is very similar to the original \eq{eq:leading_fss} and can be fitted similarly.
However, the analysis now involves three parameters: $c_0 \equiv c_G g_0$, $\omega = -y_0 / y_m$ and $y_m$.

During the completion of this work Del Debbio and Zwicky released \refcite{DelDebbio:2013qta}, where corrections to scaling due to irrelevant operators are discussed in detail, though only in infinite volume.
In the appropriate limits our results agree with \refcite{DelDebbio:2013qta}.

\section{Finite size scaling fits}
\begin{figure}[tb]
\centering
  \includegraphics[width=0.45\linewidth]{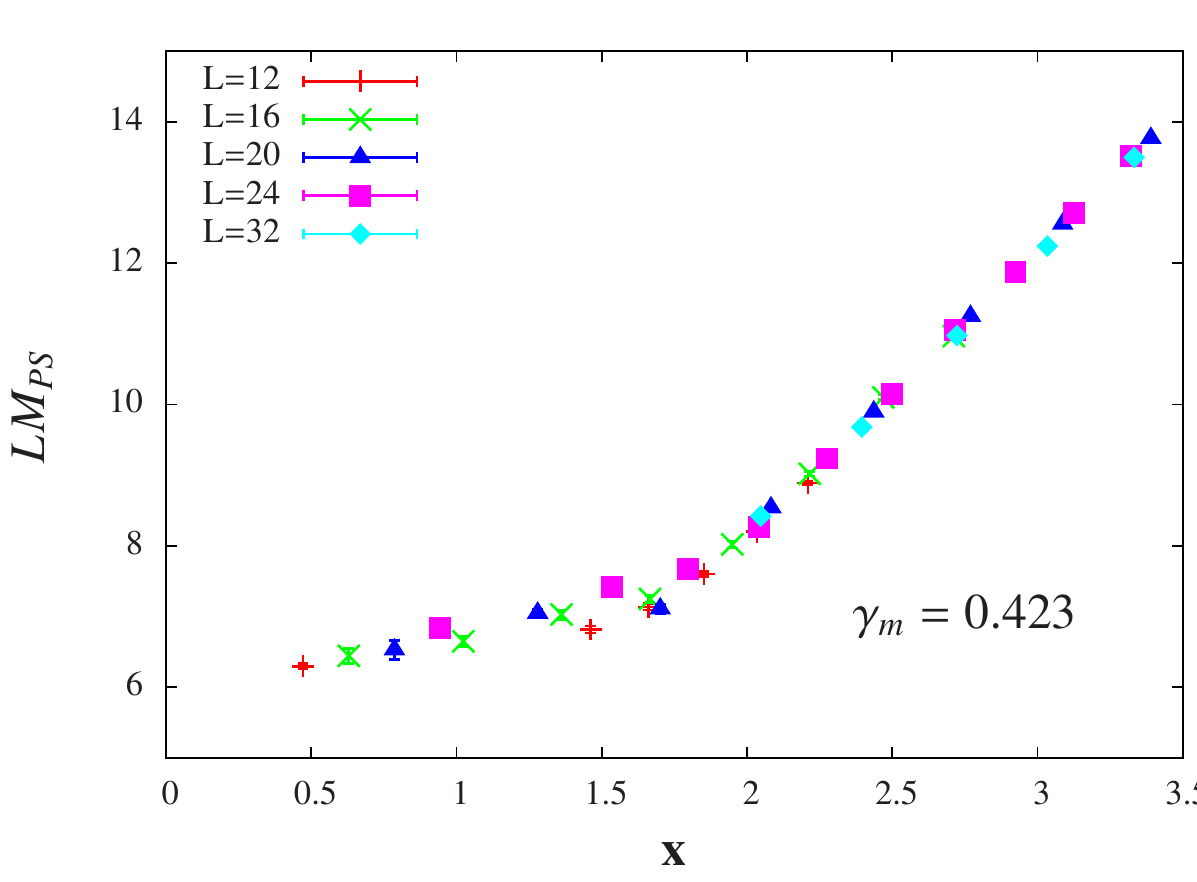}\hfill
  \includegraphics[width=0.45\linewidth]{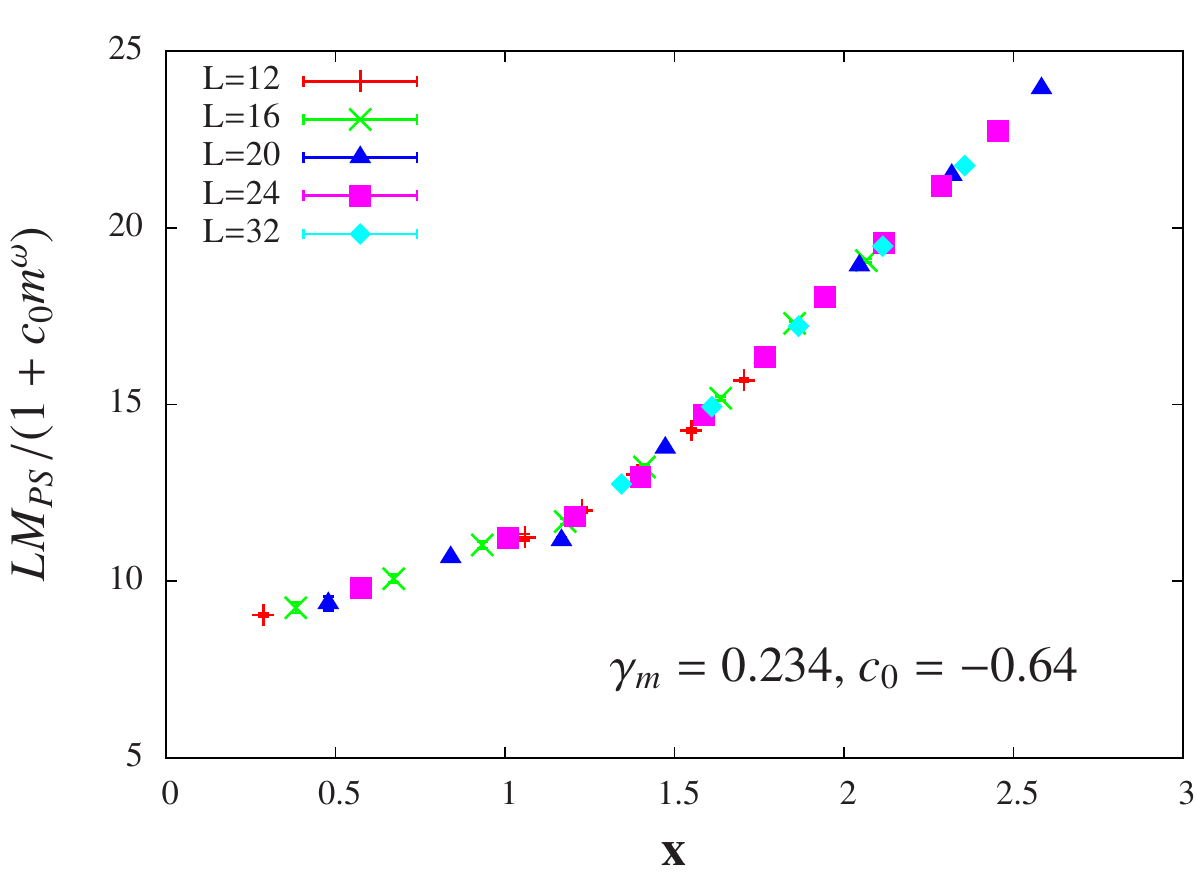}
  \caption{\label{fig:fss_pion} The best curve collapse fits for the pseudoscalar mass at $\beta_F=4.0$.  Both panels show $M_\pi L$ as the function of the scaling variable $x=m^{1/y_m} L$.  Left: Fits considering only the leading relevant operator, using Eq.~\protect\ref{eq:leading_fss}.  Right: Fits taking into account corrections, using Eq.~\protect\ref{eq:corrected}.}
\end{figure}

We begin by considering the relevant operator only, using the usual \eq{eq:leading_fss}.
The left panel of \fig{fig:fss_pion} shows the best curve collapse we found for the pseudoscalar mass at $\beta_F=4.0$.
With $y_m = 1.423$, different volumes form a unique curve for $M_\pi L \gtrsim 8$, but at smaller masses there is a clear mismatch between the different volumes (which does not improve with different values of $y_m$).
To quantify the goodness of the curve collapse we fit the data with two independent quadratic polynomials, one at $x<x_0$ and the other at $x\ge x_0$.
We minimize the $\chi^2$ of this fit in terms of $x_0$ and $y_m$.
The best fit as shown in the left panel of \fig{fig:fss_pion} has $\chi^2/\rm{dof}=6.3$.
The left panel of \fig{fig:scaling_exp} shows the results of similar analyses at other $\beta_F$ values, as well as for the vector meson and $f_\pi$.
The scaling exponents show significant variations between the three observables and as functions of $\beta_F$, suggesting that there is no consistent finite size scaling when using the form of \eq{eq:leading_fss}.

Next we take into account the leading corrections according to \eq{eq:corrected}.
We use the same two-polynomial form to fit the left side of \eq{eq:corrected} and minimize the $\chi^2$ as the function of $x_0$, $c_0$ and $y_m$ while keeping $y_0$ fixed in the range $-0.3 \lsim y_0 \lsim -0.1$~\cite{Ryttov:2010iz,Appelquist:2009ty}.
We find very little dependence on $y_0$ within this range, with slight preference for $y_0\approx -0.2$.
The right panel of \fig{fig:fss_pion} shows the best curve collapse we found for the pseudoscalar mass at $\beta_F=4.0$ using $y_0 = -0.2$.
The corresponding scaling exponent is $y_m=1.234$ with a correction term $c_0=-0.64$ and $\chi^2/\rm{dof}=3.1$.
We obtain consistent results from fitting only the small- or large-$x$ regions, justifying our approximation of constant $G(x) = c_G$. 

Repeating this analysis at other gauge couplings, and for the other two quantities considered, leads to the results in the right panel of \fig{fig:scaling_exp}, showing consistency between all three operators in the whole $\beta_F$ range investigated.
Unfortunately, the errors are significantly larger with the corrected fit, especially for $f_\pi$ where the data constrain the correction coefficient $c_0$ only weakly.
To address this we will attempt combined fits of all the data, with a universal $y_m$ and scaling functions $F(x)$ that depend on the operator but not on the gauge coupling.
The coefficients $c_0$ could depend on both the operators and the gauge coupling.
This investigation is ongoing and will be presented in a forthcoming publication.

\section{Conclusion}
We have demonstrated that apparent inconsistencies in finite size scaling analyses of the $N_f=12$ system can be resolved by considering the effect of the leading irrelevant gauge coupling, at least for $f_\pi$ and the pseudoscalar and vector meson masses.
We find that all three quantities, when considered independently, prefer an anomalous dimension $\gamma_m^{\star} = y_m^{\star} - 1 \approx 0.25$.
By performing a combined fit to all data used in this work, we hope to strengthen our conclusion and obtain a robust prediction for $\gamma_m^{\star}$.
It will also be important to consider other quantities, such as the string tension and mass of the lightest baryon, but at present we do not have these data available to analyze.

We expect that systems near the conformal boundary will generically possess a nearly-marginal gauge coupling.
The initial results presented here suggest that this may have important effects that will need to be investigated in future studies of strongly-coupled many-flavor systems.

\section*{Acknowledgments} 
We thank Julius Kuti for his helpful probing questions, Biagio Lucini for suggesting that we include corrections to finite-size scaling, and Roman Zwicky for useful discussions concerning the correction terms.
Part of this work was performed when A.~H.\ and D.~S.\ visited the Aspen Center for Physics (NSF Grant No.~1066293) and CP$^3$-Origins in Odense, and we thank both institutions for their support and hospitality.
This research was partially supported by the U.S.~Department of Energy (DOE) through Grant No.~DE-SC0010005 (A.~C., A.~H.\ and D.~S.) and by the DOE Office of Science Graduate Fellowship Program under Contract No.~DE-AC05-06OR23100 (G.~P.).
Our code is based in part on the MILC Collaboration's public lattice gauge theory software.\footnote{\texttt{http://www.physics.utah.edu/$\sim$detar/milc/}}
Numerical calculations were carried out on the HEP-TH and Janus clusters at the University of Colorado; at Fermilab under the auspices of USQCD supported by the DOE; and at the San Diego Computing Center through XSEDE supported by National Science Foundation Grant No.~OCI-1053575.

{\renewcommand{\baselinestretch}{0.86} 
  \bibliography{fss}
  \bibliographystyle{utphys}
}
\end{document}